# Answering Top-k Queries Over a Mixture of Attractive and Repulsive Dimensions

Sayan Ranu
Department of Computer Science
University of California
Santa Barbara, 93106
sayan@cs.ucsb.edu

Ambuj K. Singh
Department of Computer Science
University of California
Santa Barbara, 93106
ambuj@cs.ucsb.edu

## ABSTRACT

In this paper, we formulate a top-$k$ query that compares objects in a database to a user-provided query object on a novel scoring function. The proposed scoring function combines the idea of *attractive* and *repulsive* dimensions into a general framework to overcome the weakness of traditional distance or similarity measures. We study the properties of the proposed class of scoring functions and develop efficient and scalable index structures that index the isolines of the function. We demonstrate various scenarios where the query finds application. Empirical evaluation demonstrates a performance gain of one to two orders of magnitude on querying time over existing state-of-the-art top-$k$ techniques. Further, a qualitative analysis is performed on a real dataset to highlight the potential of the proposed query in discovering hidden data characteristics.

## 1. INTRODUCTION

Top-$k$ queries play a critical role in various applications such as business intelligence analysis, e-commerce, and virtual screening of molecular libraries. Typically, datasets for such tasks contain a large number of multidimensional objects. A top-$k$ query on such a dataset returns a subset of objects that maximize a particular scoring function. A number of techniques exist that aim to answer top-$k$ queries efficiently [3,7–10,12,18,20]. However, most of them assume a monotonic scoring function. A function is called *monotonic* if $f(x_1,..,x_n) \leq f(x'_1,..,x'_n)$, whenever $x_i \leq x'_i \; \forall i$. For example, $f(x,y) = x + y$ is a monotonic function, whereas $f(x,y) = x - y$ is not. Often situations are encountered where a monotonic function is not enough to identify the interesting objects. In this paper, we study a class of non-monotonic scoring functions that models a mixture of similarity on *attractive* dimensions and distance on *repulsive* dimensions to overcome the limitations of traditional similarity or distance measures.

Consider the online advertising scenario, where an advertiser places its advertisements in publishers' pages such as Yahoo! or Google. In such an application, an advertiser is interested in two sets of information: the cost of advertising, and the hit rate on their advertisements. Typically, top publishers charge more since an advertisement placed in their page is likely to get more hits. From an advertiser's perspective however, a high hit rate is attractive, whereas a high cost is repulsive. Thus, it is of interest to find those publishers that receive hit rates similar to that of a top publisher, and yet are cheaper.

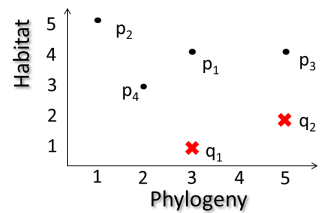

**Figure 1: A sample database and query points**

Top-$k$ queries of this nature also find application in scaffold hopping in the field of chemoinformatics. Scaffold hopping attempts to discover molecules that are structurally diverse from a given query molecule, but show similar binding activity. In chemical libraries, molecules are represented as high dimensional points in the virtual space [1, 5, 13, 14]. Further, the binding activity of a molecule against various targets is maintained as a feature vector. For scaffold hopping, given a query molecule, one needs to formulate a function that aggregates similarity on dimensions representing binding activity and distance on dimensions representing the structure.

To formalize the idea further, consider Figure 1 that presents a sample zoological database where each point represents a species. In the study of species evolution, it is often of interest to zoologists to find similar species evolving in different regions of the world. Such a query can be answered if the scoring function computes similarity on phylogeny and distance on the evolutionary habitat. Thus, given a query species, the scoring function should return database points that are as distant as possible in habitat from the query and as similar as possible in phylogeny. Therefore, for $q_1$, the desired top-1 answer is $p_1$ since its phylogeny is same as $q_1$, but the habitat is vastly different. Although both $p_2$ and $p_3$ reside in vastly dissimilar habitats, their phylogeny is distant from the query as well. Following the same reasoning, for $q_2$, the top-1 result is $p_3$; no other species is as similar in its phylogeny and as distant in its habitat.

The above function can be modeled in two ways. Assume $p$ is the absolute distance in phylogeny between a data point and the query species, while $h$ is the absolute distance in habitat between them. Thus, the top-$k$ most interesting data points can be computed using the function

$$\text{Score(data point, query)} = h - p \qquad (1)$$

An alternative approach to model the same phenomenon is to take the ratio $\frac{h}{p}$ between habitat and phylogeny. While the two functions do not behave in the exact same manner, they model the desired features. More importantly, they both share the non-monotonic behavior. The obvious question that arises from the above analysis is: *Can a monotonic function be used that models the same phenomenon and leverages existing top-k techniques?*

The non-monotonicity of the ranking function stems from the need to incorporate a query object in the ranking procedure. Most of the existing top-$k$ querying techniques assume a monotonic ranking function and typically do not incorporate a user-provided query object. In our case, the user provided query-object is central to the problem, and the entire analysis is based on the distances in the individual dimensions between the query and the database points. Since computing distance is non-monotonic, the ranking functions inherit the non-monotonicity property as well. Certainly, alternative formulations of the problem aimed at modeling similar properties are possible. However, such formulations are bound to be non-monotonic as well due to the inherent need to capture the distances between the query and the database points.

$$\text{Score}(p,q) = (c - |p_y - q_y|) + |p_x - q_x| \quad (2)$$

Consider Eqn. 2 for example. Assume $y$ is the repulsive dimension and $x$ is the attractive dimension. For query point $q = (q_x, q_y)$ and a database point $p = (p_x, p_y)$, the score can be computed by aggregating the distance in the attractive dimension and the difference between a large constant $c$, and the distance in the repulsive dimension. In this formulation, a lower score indicates a higher rank. Even though the individual contributions from the dimensions are summed, since the distance computation is non-monotonic, the formulation inherits the non-monotonicity property as well.

The above discussion shows why a monotonic function is not suitable for the proposed problem of modeling attractive and repulsive dimensions. Since computing the distance between a query and a data point on any dimension is non-monotonic, the ranking function that aggregates the individual distances is non-monotonic as well. Owing to this property, existing top-$k$ techniques that assume monotonic scoring functions are unable to solve the problem.

Among existing top-$k$ techniques, only [15] and [19] have looked at non-monotonic functions. Both techniques propose general purpose pruning strategies on hierarchical index structures. Consequently, [15] and [19] are applicable to a wide range of functions. In our solution, we take a more direct approach and develop pre-computation based index structures specifically for the proposed class of linear scoring functions. Furthermore, both [15] and [19] assume disk-resident index structures. Consequently, the techniques focus on minimizing disk accesses and guaranteeing I/O optimality. Our technique assumes a main-memory setting due to the impressive growth rate of main-memory capacities and the advent of solid state drives where random accesses are far cheaper than traditional hard disks.

In this paper, we formulate a query, called *SD-Query*, that aggregates similarity and distance into a single function. Two index structures are developed to answer top-1 and top-$k$ queries on non-monotonic scoring functions. The top-1 index structure allows us to take advantage of the fact that $k$ is known *apriori*. For example, if a data analysis program is required to return only the best candidate, then the top-1 index structure fits the application better. The second index structure is developed for the generic setting where the value of $k$ is supplied at runtime. The major contributions of the paper are summarized as follows:

**1. Query formulation**: We formulate the problem of top-$k$ query over a mixture of attractive and repulsive dimensions and demonstrate its usefulness in a number of applications.

**2. Index Structures**: We develop two novel index structures to answer top-1 and top-$k$ queries on a non-monotonic scoring function. Theoretical bounds guarantee a linear growth rate for the index structures. Empirical results demonstrate a better performance by one to two orders of magnitude over existing techniques.

## 2. PROBLEM FORMULATION

In this section, we formally define the problem and introduce the concepts that are central to the techniques developed in the paper.

**DEFINITION 1. SD-QUERY.** *Assume, we have a dataset $\mathbb{P}$ of multidimensional points $p_i=[p_{i_1},..,p_{i_m}]$. Given query $q=[q_1,..,q_m]$, sets of dimensions $\mathbb{D}$ and $\mathbb{S}$ associated with sets of weight parameters $\alpha$ and $\beta$ respectively, and an integer k, the goal is to find the k highest scoring points on the following function:*

$$\textit{SD-score}(p,q) = \sum_i \alpha_i |p_i - q_i| - \sum_j \beta_j |p_j - q_j| \; \forall i \in \mathbb{D}, \forall j \in \mathbb{S}$$
(3)

As can be seen, the above function is repulsive towards dimensions in $\mathbb{D}$ and attractive towards dimensions in $\mathbb{S}$. $\alpha_i$ and $\beta_j$ act as *weighting parameters* to tune the relative importance of the dimensions. The first summation in Eqn. 3 computes the Manhattan distance over repulsive dimensions that are desired to be as distant as possible, whereas the second summation computes the total distance over attractive dimensions chosen to be similar. As a result, SDscore is directly proportional to the distances among repulsive dimensions and inversely proportional to the distances among attractive dimensions.

To answer SD-Queries efficiently, the challenge is to develop pruning strategies that produce the answer set without scanning the entire dataset. We address that challenge by identifying the *isolines* of Eqn. 3 in the 2D plane and then designing a deterministic approach.

**DEFINITION 2. ISOLINE.** *An isoline is defined as a line that connects points of equal value in the plane for a given query.*

For dimensions above two, we divide the query into subproblems of two dimensions, and then aggregate each of the subproblems to produce the answer set. A number of existing top-$k$ querying techniques [7,8,19] take a similar approach of solving subproblems optimally and then aggregating them to compute the final answer set. However, in these techniques, a subproblem consists of a single dimension, whereas in our approach a subproblem comprises of two dimensions. As a result, a better scalability against dimension is achieved. This result is verified in Section 6.

Hereon, we assume the points to be 2-dimensional and develop pruning techniques for points in the $(x, y)$ plane. We solve the problem for arbitrary dimensions in Section 5 by extending the techniques developed in Sections 3 and 4. Without loss of generality, we consider dimension $x$ for similarity and $y$ for distance. Therefore, for query $q = [x_q, y_q]$, we try to maximize the following function:

$$\text{SD-score}(p,q) = \alpha |y_p - y_q| - \beta |x_p - x_q| \; \forall p \in \mathbb{P} \quad (4)$$

EXAMPLE: *If $\alpha = \beta = 1$, for the sample database in Figure 1, $SDscore(p_1, q_1) = 3 - 0 = 3$ and $SDscore(p_3, q_2) = 2 - 0 = 2$.*

Consider the example shown in Figure 2a, which shows points from a sample dataset and a given query point $q$. Assume both dimensions are equally important resulting in $\alpha = \beta = 1$. Given

170

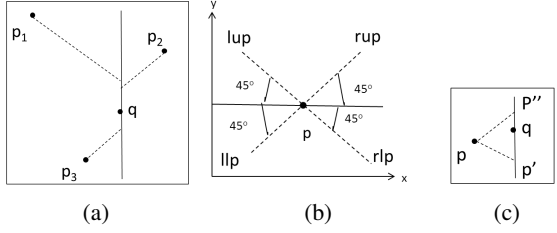

(a) (b) (c)

**Figure 2:** (a) $45°$ degree projections of points on query $q$. (b) Illustration of the four possible projections from a point. (c) Illustration of the scenario when SD-score$(p,q) \leq 0$

such a scenario, the solid line across $q$ represents its *axis* and the dashed lines emerging from the database points are termed as *projections*. The projections are lines originating from data points $p_i$s at $45°$ angle. Due to this geometry, they have a special property with respect to $q$. For each point, the score with respect to $q$ is the same as the score between $q$ and the intersection point of its projection with $q$'s axis since these points lie on the same *isoline*. For any arbitrary $\alpha$ and $\beta$, the angle can be computed as:

$$\theta = \arctan \frac{\beta}{\alpha} \qquad (5)$$

**DEFINITION 3.** AXIS. *The axis of a query point $q = [x_q, y_q]$ is defined by the line $x = x_q$.*

**DEFINITION 4.** PROJECTION. *Projection of a point $p$ is a line originating from $p$ at an angle $\theta$ computed using Eqn. 5. In the 2D plane, each point generates four projections in four directions. We call these projections left lower projection (abbr. llp), right lower projection (abbr. rlp), left upper projection (abbr. lup) and right upper projection (abbr. rup). An example is shown in Figure 2b. From geometric reasoning, projections of the same type (such as llp) are parallel to each other.*

EXAMPLE: *In Figure 2a, the three shown projections are of types rlp ($p_1$), llp ($p_2$) and rup ($p_3$).*

Hereon, for explanatory purposes and without loss of generality, we assume $\alpha = \beta = 1$ for simplicity resulting in $\theta = 45°$. All the theory developed in this paper holds for any arbitrary $\alpha$ and $\beta$.

Note, in Figure 2a, only one of the four projections is shown for each point. However, as can be seen in Figure 2b, four projections exist for each point, and only one of them is the correct isoline with respect to a query. Thus, given query $q$, it is important to deterministically choose the correct projection for a point $p$. Towards that goal, we make the following observations. First, if $x_q < x_p$, we only need to decide between the left projections of $p$ since the right projections will never intersect $q$'s axis. Second, if $y_p < y_q$, then any of the lower projections will not provide the correct score. Analogous relationships exist for $x_q > x_p$ and $y_p > y_q$. Based on these observations, we make the following claims:

**CLAIM 1.** *For query $q=[x_q, y_q]$, if projections from $p=[x_p, y_p]$ intersect $q$'s axis at points $p''$ and $p'$, and $q$ is located between $p''$ and $p'$, then SD-score$(p, q) \leq 0$. An example is shown in Figure 2c.*

PROOF: Due to geometric constraints,
$p'' = [x_q, y_p + |x_q - x_p|]$
$p' = [x_q, y_p - |x_q - x_p|]$
Thus, $y_q = y_p - |x_q - x_p| + c$ where $0 \leq c \leq 2|x_q - x_p|$.

SD-score$(p, q) = |y_p - y_p + |x_q - x_p| - c| - |x_p - x_q|$
$= \begin{cases} -c, & \text{if } (|x_p - x_q| - c) \geq 0 \\ c - 2|x_p - x_q|, & \text{if } (|x_p - x_q| - c) < 0 \end{cases}$
$\leq 0$ □

Now, let us define a function $Projection(p, q)$ for points $p=[x_p, y_p]$ and query $q=[x_q, y_q]$.

$$\text{Projection}(p, q) = \begin{cases} lup, & \text{if } y_p < y_q \text{ and } x_p \geq x_q \\ llp, & \text{if } y_p \geq y_q \text{ and } x_p \geq x_q \\ rup, & \text{if } y_p < y_q \text{ and } x_p < x_q \\ rlp, & \text{if } y_p \geq y_q \text{ and } x_p < x_q \end{cases} \qquad (6)$$

Hereon, whenever we refer to a projection of a point $p$ on a query, we assume it to be the projection chosen according to Eqn. 6. Moreover, we refer to the intersection point between $p$'s projection and $q$'s axis as $p$'s *projected point* on $q$.

Let the intersection point between Projection$(p, q)$ and $q$'s axis be $p'=[x_q, y_{p'}]$.

**CLAIM 2.** *If $p$ does not satisfy the conditions in Claim 1, then SD-score$(p, q)$=SD-score$(p', q)$.*

PROOF: There are two possible cases. $y'_p \geq y_q$ and $y'_p < y_q$.
**For Case 1:** $\qquad y'_p \geq y_q \qquad (7)$

Since projections follow a $45°$ angle and $p$ does not satisfy the conditions in Claim 1,

$$y_{p'} = y_p - |x_q - x_p| \qquad (8)$$

SD-score$(p', q) = |y_p - |x_q - x_p| - y_q|$
$= \begin{cases} |y_p - y_q| - |x_q - x_p|, & \text{if, } y_p - y_q \geq |x_q - x_p| \\ |x_q - x_p| + y_q - y_p, & \text{if, } y_p - y_q < |x_q - x_p| \end{cases}$
$= \text{SD-score}(p,q)$

since $y_p - y_q < |x_q - x_p|$ contradicts Eqn. 7.

**Case 2:** The proof follows analogously. □

**CLAIM 3.** *If $p$ satisfies conditions in Claim 1, then SD-score$(p, q) = -|y_q - y_{p'}|$*

PROOF: There are two possible cases. $y_p \geq y_q$ and $y_p < y_q$.
**For Case 1:** $y_p \geq y_q$.
Since a lower projection (rlp or llp) would be selected,

$y_{p'} = y_p - |x_q - x_p|$
SD-score$(p, q) = |y_q - y_p| - |x_q - x_p|$
$= y_p - y_q - |x_q - x_p|$
$= -|y_{p'} - y_q|$

**Case 2:** The proof follows analogously.[1] □

Claim 1 formalizes the conditions under which a point is guaranteed to return a negative score. Claim 2 establishes the projections as the *isolines* of all data points with positive scores. For the remaining points with negative scores, although their projections are not the isolines for a given query, Claim 3 shows that the scores can still be computed using just the projections. Next, we show that to compute the top-$k$ answer set for any given query, the search can be limited to only those points that correspond to either the top-$k$ highest lower projections or the top-$k$ lowest upper projections on the query's axis.

---
[1]For arbitrary $\alpha$ and $\beta$, SD-score$(p, q) = -\alpha|y_{p'} - y_q|$

CLAIM 4. *For any query q, the top-k result is a subset of the points corresponding to the highest lower projections and the points corresponding to the lowest upper projections on q's axis.*

PROOF: With respect to $q$, any data point $p$ can be divided into one of the two groups: $y_p \geq y_q$, and $y_p < y_q$.
**Group 1:** For the group $y_p \geq y_q$, Projection$(p, q)$ is always a lower projection (i.e. rlp or llp). Now, based on whether $p$ satisfies Claim 1, Group 1 can further be divided into two subgroups: $y_{p'} \geq y_q$, and $y_{p'} < y_q$, where $p'$ is $p$'s projected point on $q$.
*Subgroup 1:* $y_{p'} \geq y_q$. From Claim 2, SD-score$(p, q) = y_{p'} - y_q$.
*Subgroup 2:* $y_{p'} < y_q$. From Claim 3, SD-score$(p, q) = y_{p'} - y_q$.
Thus, for both subgroups, SD-score$(p, q)$ increases with $y_{p'}$. As a result, the top-$k$ answer set within this group is equivalent to the points with the top-$k$ highest lower projections on $q$.
**Group 2:** For the group $y_p < y_q$, Projection$(p, q)$ is always an upper projection. It can be analogously shown that SD-score$(p, q)$ decreases with $y_{p'}$ as a result of which, the top-$k$ answer set is equivalent to the points with the lowest upper projections.
Thus, under all situations, the top-$k$ answer set is a subset of the points corresponding to the top-$k$ highest lower projections or the top-$k$ lowest upper projections. □

EXAMPLE: *In Figure 2a, the highest lower projection and lowest upper projection on q come from $p_1$ (rlp) and $p_3$ (rup) respectively. Thus, if we are looking to compute the top-1, a comparison is required only between points $p_1$ and $p_3$.*

Claim 4 establishes that the search space for the top-$k$ answer set can be pruned drastically by analyzing the projections on the given query's axis. However, projecting the database points on the query's axis, and then determining the top-$k$ projections keep the cost linear. In the next two sections, we thus focus on how the discovered properties can be used to compute the top-$k$ answer set in sublinear time.

## 3. INDEX STRUCTURE FOR TOP-1

In this section, we develop an index structure under the assumption that $k$ and the weighting parameters $\alpha$ and $\beta$ are known apriori. Although we make the assumption that $k = \alpha = \beta = 1$, the techniques developed in this section are generalizable to arbitrary values of $k$, $\alpha$, and $\beta$. The apriori knowledge of the parameters is used to design a highly compact and efficient index structure. The advantage of the apriori knowledge of the parameters in reducing storage and computation costs are quantified in Section 6. We generalize the index structure for the case where the parameters are supplied at query time in the next section.

Given the knowledge that $k = 1$, using Claim 4, a point $p$ is a candidate for top-1 only if its projections on query $q$'s axis is either the lowest upper projection or the highest lower projection. Although this result allows us to prune majority of the data points, the overhead of first projecting all data points on $q$'s axis keeps the cost linear. Thus, we ask the question: *can the two extreme projections on q be preserved and retrieved in a scalable manner?*

To answer the question, we analyze the scenario shown in Figure 3. For simplicity, we consider just the lower projections. Consider query $q_2$. The highest lower projection on its axis is from $p_1$. However, for query $q_1$, the highest lower projection is from $p_2$. The result shows that the ordering between the lower projections has changed due to the shift in the location of the two queries. It can be seen that the ordering of a projection can change only if it intersects with another one. In this particular case, the intersection is between the rlp of $p_2$ and llp of $p_1$. Further, by Claim 4 and due to geometric constraints, the highest projection on any query $q$ can change only through an intersection between the rlp and llp of two points, while the lowest projection can change due to an intersection between rup and lup.

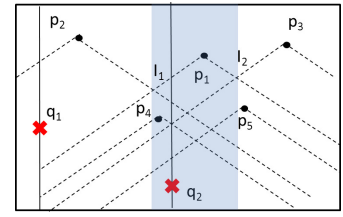

**Figure 3: Illustration of how the ordering between projections changes due to intersections. For simplicity, only the lower projections are shown.**

EXAMPLE: *Figure 3 provides a detailed illustration of how the ordering between projections changes. As shown in the figure, $p_1$ provides the highest projection in the shaded region between $I_1$ and $I_2$ for any given query. At $I_1$, the rlp of $p_2$ and llp of $p_1$ intersect due to which the ordering changes. Thus, $p_2$ provides the highest projection in the region left of $I_1$. Moreover, $p_1$ will never provide the highest projection either to the left or right of the shaded region due to the geometry of the projections.*

Based on the observations noted in the above example, we make the following claim:

CLAIM 5. *For any point p, there can exist at most one continuous region r, where p provides the highest lower projection. The statement also holds for lowest upper projections.*

PROOF: **Case 1:** Let us first consider the case of highest lower projections. Assume $p$ provides the highest lower projection in region $r$. The boundaries of $r$ are defined by vertical lines running through the points at which the llp or rlp of $p$ is intersected by a projection from some other point. Now, for $p$ to provide the highest lower projection again in a region outside $r$, one of its lower projections must become the highest again. However, we show that it is never possible.
*Subcase 1.1: Right lower projection of p becomes the highest projection again.* Assume the llp of some point $\hat{p}$ intersects the rlp of $p$. To the right of the intersection point, the rlp of $p$ can never again be the highest lower projection since:
1. throughout its extent, llp of $\hat{p}$ remains above rlp of $p$.
2. when llp of $\hat{p}$ ends at $\hat{p}$, rlp of $\hat{p}$ starts and remains above rlp of $p$ throughout.
The proofs of Subcase 1.2 for llp of $p$ and Case 2 for lowest upper projections follow analogously. □

From Claim 5, it follows that in a dataset of $n$ points, there can be at most $n$ unique regions corresponding to points providing the highest lower projections and $n$ more for lowest upper projections. The intersection of these regions results in at most $2n$ regions of unique pairs of highest and lowest projection providers. This result establishes that for any dataset, the 2D space can be divided into regions, such that in each region the two points corresponding to the highest and lowest projections remain static. Now, given any query $q$, the highest and lowest projections on its axis can be retrieved by finding the region in which $q$'s axis lies. Once the region is found, computing the answer takes constant time since it involves comparing the points corresponding to the two extreme projections.

Motivated by the above observation, we focus on developing an algorithm to efficiently index regions with unique pairs of highest and lowest projection providers. First, we scan the space left

## Algorithm 1 ConstructTop-1Index($\mathbb{P}$)

**Require:** $\mathbb{P}$ is a set of points
1: $P_{llp} \leftarrow$ points sorted in descending order of its llp on $x = -\infty$
2: $P_{lup} \leftarrow$ points sorted in ascending order of its lup on $x = -\infty$
3: $TopLlp \leftarrow P_{llp}[1]$
4: $TopLup \leftarrow P_{lup}[1]$
5: $upperIndex \leftarrow$ empty array
6: $lowerIndex \leftarrow$ empty array
7: **for** $i$ in $range(2, |\mathbb{P}|)$ **do**
8:     $next \leftarrow P_{llp}[i]$
9:     **if** $llp$ of $next$ intersects $rlp$ of $TopLlp$ **then**
10:         $in \leftarrow$ intersection between rlp of $TopLlp$ and llp of $next$
11:         add($upperIndex, \langle TopLlp, x_{in} \rangle$)
12:         $TopLlp \leftarrow nextTopLlp$
13:     $next \leftarrow P_{lup}[i]$
14:     **if** $lup$ of $next$ intersects $rup$ of $TopLup$ **then**
15:         $in \leftarrow$ intersection between rup of $TopLup$ and lup of $next$
16:         add($lowerIndex, \langle TopLup, x_{in} \rangle$)
17:         $TopLup \leftarrow nextTopLup$
18: add($upperIndex, \langle TopLlp, \infty \rangle$)
19: add($lowerIndex, \langle TopLlp, \infty \rangle$)
20: **return** merge($upperIndex, lowerIndex$)

to right to identify points that correspond to a highest lower projection, or a lowest upper projection within some region $r$. When such a point is found, the point and the boundary of region $r$ are stored in the index structure. The resultant index structure can be viewed as a sorted array of regions. The first cell in the array corresponds to the leftmost region, whereas the last cell corresponds to the rightmost region. The boundary of a region is identified by the *x-intercept* of the vertical line that passes through the intersection of the projections between two consecutive regions. For higher values of $k$, instead of tracking the highest and lowest projections, we need to track the $k$-highest and $k$-lowest projections. Further, any region where the ordering of the $k$-highest or $k$-lowest projections changes, needs to indexed.

Alg. 1 presents the pseudocode for the index construction algorithm. First, two sorted lists are constructed to facilitate the line-sweep algorithm. The first list arranges points in descending order of the llps when projected on the line $x = -\infty$ (line 1). As a result, the first point in the list corresponds to the highest llp incident on $x = -\infty$. Similarly, the second list is ordered based on lups in ascending order (line 2). From each of the ordered lists, the top elements are fetched (lines 3-4). Both top elements are guaranteed to be in the index, since they correspond to either the highest or lowest projection. In each of the next iterations, the next points in the lists are fetched. For the upper index, the next point is added if its llp intersects with rlp of the current top element. Along with the point, the *x-value* of the intersection point is added. The vertical line through the intersection point defines the region boundary. Similar checks are made among upper projections to add points to the lower index (lines 7-17). Once the iteration completes, the upper and lower index structures are merged. The merging procedure is same as merging two sorted arrays. Finally, the merged index structure is returned (line 20).

EXAMPLE: *Consider Figure 3 as our sample database. For simplicity, we consider only the highest lower projections. Alg. 1 first creates a sorted list based on the llps of all points on the line $x = -\infty$. The first point in the list is $p_2$ since it provides the highest projection. To determine the extent of the region where $p_2$ provides the highest lower projection, the second point in the list, $p_1$, is fetched to check if it intersects the rlp of $p_2$. Since the check is true for $p_1$, a cell is created for $p_2$ which stores the point itself and the x-value of the intersection point $I_1$. The next point fetched from the sorted list is $p_4$. However, $p_4$ is discarded since its llp does not intersect the rlp of $p_1$. Continuing in the same manner, since $p_3$'s llp intersects the rlp of $p_1$, a second cell containing point $p_1$ and the x-value of the intersection point $I_2$ is added. Finally, a cell is also created for $p_3$ with a x-value of $\infty$ since its rlp is not intersected by any of the database points. Thus, the algorithm produces an array of three cells corresponding to points $p_2$, $p_1$, and $p_3$. Both $p_4$ and $p_5$ get discarded since they never provide the highest lower projection.*

As can be seen, the index construction algorithm keeps the regions sorted based on the *x-intercepts* of the boundaries. Now, given any query $q=[x_q, y_q]$, the region $q$'s axis lies in can be found by performing a binary search $x_q$ on the index. Once the region is found, a comparison between the two points corresponding to the highest and lowest projections in that region produces the answer.

The above algorithm indexes only those points that have a chance of being in the top-1 answer set for any given query. Below, we discuss the storage and computation complexities of the proposed approach for a database with $n$ points.

**1. Storage Cost:** There can be at most $2n$ regions. For each region, the two points corresponding to the highest and lowest projections, and the *x-intercept* of the boundary are stored. This results in $O(n)$ storage cost.

**2. Querying Cost:** For each query, a binary search is performed with the *x-intercept* of its axis. The binary search costs $O(\log n)$ time. Once the region is found, computation of top-1 requires constant time. Thus, the total cost is $O(\log n)$.

**3. Index Construction Time:** To construct the index structure, projections of each point needs to be sorted. Once sorted a modified version of the line-sweep algorithm is performed to find the regions in linear time. Thus, the total cost is $O(n \log n)$.

When generalized to top-$k$, the storage cost, querying cost, and index construction time are $O(kn)$, $O(\log n + k)$, and $O(n \log n + nk)$ respectively.

Next, we discuss how updates can be made on an existing index.
**1. Insert:** To insert a point $p$, a binary search is performed to identify the region where $p$ lies. If $p$ does not provide the highest lower projection and lowest upper projection in that region, $p$ does not need to be inserted. Otherwise, the left projections of $p$ are compared iteratively to the indexed points with *x-values* less than that of $p$ and the right projections are compared to the indexed points with higher *x-values*. The iteration continues till a left projection of $p$ is intersected by the right projection of an indexed point and similarly the right projection of $p$ is intersected by the left projection of an indexed point. The intersections define the region within which $p$ provides the highest or lowest projections. The computation cost of the operation is bounded by $O(n)$.
**2. Delete:** Since the top-1 index only stores points that have a chance of being in the answer set, no changes are necessary for deleting a data point $p$ that is not in the index. Otherwise, the index is rebuilt only within the region where $p$ provides the highest or lowest projection using the same method as in Alg. 1. Note, we do not need to recompute or sort the projections of the data points since they were already computed while constructing the index. Thus, deleting a point incurs $O(n)$ cost.

## 4. INDEX STRUCTURE FOR TOP-K

In this section, we develop an index structure for top-$k$ queries. A straightforward extension of the structure developed for top-1 does not work since $k$ and the weighting parameters are supplied at runtime. Specifically, there are two main bottlenecks. First, if we are to extend the same technique, for each region, we need to store the entire ordering of projections, thus requiring $O(n^2)$ storage for the entire index structure. Second, since the weighting parameters are supplied only at query time, a static ordering among the projections cannot be assumed. The challenge is therefore to keep the storage requirement linear and answer top-$k$ queries at logarithmic computation cost in spite of the added complexities. We tackle this challenge by dividing the problem into two subproblems: first, we develop an index structure assuming fixed angles of projections. Next, we adapt the index structure for the general case where the weighting parameters are provided at query time.

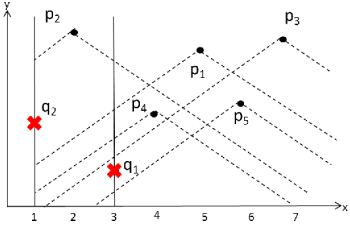

**Figure 4: A sample database of points.** $p_i$ represents database points and $q_i$ represent query points

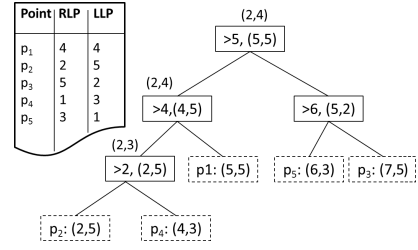

**Figure 5: Index structure on points in Figure 4 for top-$k$ query.** Each non-leaf node contains the *x-intercept* of the separating plane and the tuple representing the bounds on rlp and llp in its subtree. The tuples shown on top of the nodes represent the bounds after the update operation for query $q_1$. The leaves contain the coordinates of the points.

### 4.1 Fixed angle of projection

Given fixed angle of projections, we observe that due to geometric constraints, projections of the same type are parallel to each other. For example, the rlp of all points are parallel to each other. As a result, projections of the same type never intersect among themselves, and thus maintain a fixed ordering. Thus, projections can be presorted into four lists based on their type. Now, given query $q$, if we are able to identify the top-$k$ highest rlp, llp, and top-$k$ lowest rup, lup on $q$'s axis, then the search for the top-$k$ points can be limited to the $4k$ points corresponding to the identified projections (Claim 4). However, retrieving the top-$k$ projections on $q$ is not an easy problem since projections cover only part of the 2D space. An example is shown in Figure 4 to illustrate the point. For simplicity, we only analyze the highest lower projections on $q$, and thus limit ourselves to the rlp and llp of the points. For query $q_1$, five projections intersect its axis: llp from $p_1$, $p_3$, $p_4$, and $p_5$ and rlp from $p_2$. On the other hand, for $q_2$, the llp from $p_2$ intersects its axis while the rlp does not. It can be seen, that for an llp or lup to intersect the query's axis, the corresponding point must be located to the right of the query. Analogously, points corresponding to rup and rlp must be located to the left of the query. Thus, to solve the problem of finding the top-$k$ projections on a query, we need to perform a range search to retrieve the projections that intersect the axis. However, instead of returning all projections in the range, the search should only return the top-$k$ projections.

Towards that goal, we develop a tree that facilitates both range search and top-$k$ queries. First, a tree is built on the *x-values* of points to facilitate range search. The tree is constructed in a manner similar to KD-tree [6], but on a single dimension with a branching factor $b$. At each non-leaf node, the dataset is divided in a balanced manner into $b$ subsets and the process is repeated recursively. Each subset contains points with *x-values* less than or equal to a chosen separating plane. For example, if the branching factor is 2, the median of the points is chosen as the separating plane. The *x-intercepts* of the separating planes are stored at each non-leaf node. All points are stored at leaf nodes.

Once the tree is constructed, for each point in the dataset, its llp and lup are projected on the line $x = -\infty$, while the rup and rlp are projected on the line $x = \infty$. Projecting on the two extreme lines allows us to order all projections in the dataset. Next, at each non-leaf node, we store the bounds on the highest and lowest projections in its subtree. Specifically, the *y-value* of the intersection point between the highest rlp, llp, and the lowest rup, lup with the relevant line, which is either $x = -\infty$ or $x = \infty$, is calculated and stored at the non-leaf nodes. With this information, the tree inherits a heap-like property.

EXAMPLE: *Figure 5 presents the tree built with a branching factor of 2 on points shown in Figure 4. For simplicity, the tree tracks only rlp and llp. The y-values of the intersection points of the projections with the line $x = -\infty$ and $x = \infty$ are also shown in Figure 5. Without loss of generality, the y-values are chosen arbitrarily, but maintain the same ordering among projections. The leaves are represented as dashed nodes and contain the points. Each of the non-leaf nodes contains the x-intercept of the separating plane, and a tuple representing the highest rlp and llp in its subtree.*

The constructed tree allows fast answering of queries. The pseudocode of the algorithm is provided in Alg. 2. Given query $q$, a range search is performed to group the points into two sets based on which side of $q$'s axis they lie on. Specifically, at each node, the left child is chosen if $x_q$ is less than or equal to the *x-intercept* of the separating plane, otherwise, the right child is chosen. For higher branching factors, the traversal is generalized in the same manner. The search stops at a leaf node and the path from the root to this node acts as the dividing plane for the points. We call this the *"separating path"*.

EXAMPLE: *A range search using $q_1$ (shown in Figure 4) on the index in Figure 5 ends at $p_4$, and the path from the root to $p_4$ is the separating path. The llp and lup from all nodes in and right of the separating path intersect $q$'s axis. Similarly, rlp and rup from nodes left of the separating path intersect as well.*

---

**Algorithm 2** top-k query($Index, q, k$)

**Require:** $q$ is a query point
1: $answer \leftarrow$ empty array of size $k$
2: $cand \leftarrow$ empty array of size 4
3: $r \leftarrow$ root of $index$
4: updateBounds($r, q$)
5: insert getTop($q$,llp), getTop($q$,lup), getTop($q$,rlp), getTop($q$,rup) in $cand$
6: **while** $size(answer) < k$ **do**
7: $\langle p, type \rangle \leftarrow$ highest scorer in $cand$ and corresponding projection type
8: add $p$ to $answer$
9: insert getTop($q$,$type$) to $cand$
10: **return** answer

**Algorithm 3** updateBounds($n, q$)

**Ensure:** finds *separating path* and updates bounds on the path
1: **if** $n$ is leaf **then**
2:   **return**
3: $llpBound \leftarrow \max(llpBound(child)) | \forall child \in n, x_{child} \geq x_q$
4: $lupBound \leftarrow \min(lupBound(child)) | \forall child \in n, x_{child} \geq x_q$
5: $rlpBound \leftarrow \max(rlpBound(child)) | \forall child \in n, x_{child} \leq x_q$
6: $rupBound \leftarrow \min(rupBound(child)) | \forall child \in n, x_{child} \leq x_q$
7: $pos \leftarrow$ binarysearch(array of x-intercepts of children, $x_q$)
8: updateBounds($child[pos], q$)

Now, the goal is to get the top-$k$ of each of the intersecting projections. For that task, an update operation is performed on the *separating path*. At each node along the *separating path*, bounds on the highest llp and lowest lup are updated to the highest and lowest among the children located in or right of the path. Analogously, bounds on the right projections are updated among nodes left of the *separating path* ( Alg. 3). The updated values for $q_1$ are shown on top of each node in the *separating path* in Figure 5. With this update operation, the root contains the highest and lowest values only among projections that are incident on the query's axis. For example, even though the highest llp is from $p_2$, due to the update operation, the llp bound at root changes to 4 corresponding to $p_1$. Next, searching for the top-$k$ on a particular projection type is performed by starting at the root and traversing to the child which has the same value for that type. This traversal is performed recursively at each node till a leaf-node with the same value is reached. From the index construction algorithm, it is guaranteed that at any node, one of its children will have the same value. Thus, each search step deterministically takes us closer to the answer.

EXAMPLE: *Assume we are searching for highest llp on $q_1$ in Figure 5. The llp value at root is 4, and its left child has the same value as well. Thus, the first step traverses to the left child of the root. In the next step, the correct answer $p_1$ is selected and returned.*

Once the answer is returned, the bound in the parent node is recalculated by ignoring the current answer. Further, the change in bound is propagated upwards along the traversed path. The change in the bounds facilitates the search of the next element in top-$k$. Once the update operation is complete, the cycle for top-1 ends.

EXAMPLE: *In the update operation for $q_1$, the parent of $p_1$ is updated to $(2, 3)$ since the next highest llp in the subtree that intersects $q_1$'s axis is $p_4$. The change is then propagated to the root which gets updated to $(2, 3)$ as well.*

For the top-$k$ search, once the *separating path* is found and the bounds are updated, a top-1 search is made on each of the four projection types (Alg. 2, line 5). Among the four points returned, the highest scorer is added to the answer set and the remaining three are retained as candidates. For the next subsequent searches, the search is performed only on the projection type that contributed to the most recent point in the answer set (line 6-9). For example, if in the first iteration, the highest scorer corresponds to projection type llp, then in the next iteration the search would be performed only on llp. The algorithm terminates after $k + 3$ searches.

Below, we discuss the storage and computation complexities of the index constructed on $n$ data points and a branching factor $b$.
**1. Storage Cost:** Since the tree is constructed in a balanced manner, the height is $\log_b n$. Each level of the tree follows a geometric progression resulting in $O(\frac{nb-1}{b-1})$ storage.
**2. Querying Cost:** For any query, first the *separating path* is found at $b \log_b n$ time. After the update, $k + 3$ top-$k$ searches are performed where each search consumes $2b \log_b n$ time. Further, whenever a point is returned, four comparisons are made for insertion into the answer set. Thus, the total querying time is bounded by $O(kb \log_b n + k)$.
**3. Index Construction Time:** To construct the index structure, first the points are sorted based on their *x-values*. The $n \log n$ cost of sorting dominates other index construction operations of projecting points to $x = -\infty$ and $x = \infty$ ($O(n)$) time, and addition of bound information on each non-leaf node ($O(n \log_b n)$). Thus, the overall time complexity is $O(n \log n)$.

Next, we briefly discuss how the top-$k$ index can be adapted for a disk-resident version. Since, the shape of the proposed top-$k$ index structure is highly similar to B+-tree, similar index construction techniques can be employed. Each node can be stored in a single disk page. Further, instead of a uniform branching factor, each of the non-leaf nodes should contain between $c$ and $2c$ children, where $c$ is chosen according to the disk page size. Similar to B+-trees, for efficient construction of the index structure, *bulk-loading* can be employed. More specifically, the data can first be sorted based on their *x-values*, and the index can then be built in a bottom-up manner. For each level of the index, the nodes can be packed entirely full, except for the rightmost node. Once the tree is built, the bounds on the projections can be added. During query time, majority of the search proceeds in the same manner as in the in-memory version. However, since each of the leaf nodes contains multiple data points in the disk-resident version, a comparison among those points is required to identify the one with the highest score.

Next, we discuss how updates can be performed efficiently.
**1. Insert:** The insert operation for a new point $p$ starts at the root node and recursively collides to the appropriate child based on the *x-value*. At the end of the path, two cases are possible. (i) $p$ collides with another leaf node $l$. In this case, a new non-leaf node replaces $l$. The new non-leaf node contains both $l$ and $p$. (ii) $p$ does not collide with another leaf node. In this case, $p$ is simply inserted as a new leaf node. After $p$ is inserted, its projections are computed, and the bounds on the path to the root are updated. The insert operation takes $O(\log_b n)$ time.
**2. Delete:** A delete operation deletes the leaf-node corresponding to the chosen point and updates the projection bounds on the path from the leaf to the root. The operation takes $O(b \log_b n)$ time.

While updates can be made in an efficient manner, each update can make the tree unbalanced. More specifically, the height of the tree can exceed $\log_b n$, resulting in slower querying times. We tackle this problem by keeping track of the set of leaf-nodes, $\mathbb{U}$, on paths exceeding the length of $\log_b n$. An unbalanced index would affect the querying time if the answer set contains points from $\mathbb{U}$. Thus, when the probability of longer querying times, $\frac{|\mathbb{U}|}{n}$, exceeds a threshold $\theta$, we rebuild the index structure.

### 4.2 Answering queries with arbitrary weighting parameters

Although the top-$k$ index structure developed is efficient in answering top-$k$ queries, its applicability is significantly hampered due to the assumption of a fixed angle of projection. Thus, to remove this bottleneck, we extend the index structure to handle queries with arbitrary weighting parameters on the repulsive and attractive dimensions. As discussed in the Section 2, the angle of projection can be computed using Eqn. 5.

To make the index structure more flexible, we observe the following properties of the projections.
**1.** The angle of any projection in a quadrant lies in range $[0°, 90°]$.
**2.** Let Score($p, \theta$) represent the score of a point $p$ when projected at an angle $\theta$ on a query $q$'s axis. For two points $p_1$ and $p_2$ and angles $\theta_1, \theta_2, \theta_3$ such that $\theta_1 < \theta_2 < \theta_3$, if Score($p_1, \theta_1$)$\geq$ Score($p_2, \theta_1$) and Score($p_2, \theta_2$)$\geq$ Score($p_1, \theta_2$) then Score($p_2, \theta_3$)$\geq$ Score($p_1$,

**Algorithm 4** top-$k$ArbitraryParameters($Index, q, k, \alpha, \beta$)
─────────────────────────────────────────
1: $\theta_q \leftarrow \arctan \frac{\beta}{\alpha}$
2: Find indexed angles $\theta_l$ and $\theta_u$ where $\theta_l \leq \theta_q \leq \theta_u$
3: top-$k_{\theta_l} \leftarrow$ top-$k$ answer set on $q$ at $\theta_l$
4: top-$k_{\theta_u} \leftarrow \emptyset$
5: top-$k_{\theta_q} \leftarrow$ empty priority queue of maximum size $k$
6: **while** top-$k_{\theta_l} \not\subseteq$ top-$k_{\theta_u}$ **do**
7:    add $p$ to top-$k_{\theta_u}$
8:    evaluate score of $p$ at $\theta_q$ and add $\langle p, \text{score}(p, \theta_q) \rangle$ to top-$k_{\theta_q}$
9: **return** top-$k_{\theta_q}$
─────────────────────────────────────────

$\theta_3$). This result follows from the geometry of the projections.

Based on the above properties we claim the following.

CLAIM 6. *Let us denote the top-$k$ answer set for $q$ at any angle $\theta$ as top-$k_\theta$. Assume, we have two index structures for projections at angles $\theta_l$ and $\theta_u$ and the angle of projection for a query $q$ is $\theta_q$, where $\theta_l \leq \theta_q \leq \theta_u$. Then,*

$$\text{top-}k_{\theta_q} \subseteq \arg\min_{k'}\{\text{top-}k'_{\theta_u} \text{ s.t. top-}k_{\theta_l} \subseteq \text{top-}k'_{\theta_u}\} \quad (9)$$

PROOF BY CONTRADICTION: Assume top-$k_{\theta_q} \not\subseteq$ top-$k'_{\theta_u}$. Therefore, there is at least one point $p$, where $p \in$ top-$k_{\theta_q}$ and $p \notin$ top-$k'_{\theta_u}$, and therefore, $p \notin$ top-$k_{\theta_l}$. Thus, there is some point $p' \in$ top-$k_{\theta_l}$, where Score($p', \theta_l$)$\geq$ Score($p, \theta_l$) and Score($p, \theta_q$)$\geq$ Score($p', \theta_q$). Therefore, based on observation 2 in Section 4.2, Score($p, \theta_u$)$\geq$ Score($p', \theta_u$) which is a contradiction since, in assumption, $p' \in$ top-$k'_{\theta_u}$ and $p \notin$ top-$k'_{\theta_u}$. □

Claim 6 provides a framework to compute the top-$k$ answer set on a non-indexed angle of projection if the angle is bounded between two angles $[\theta_l, \theta_u]$ that are already indexed. First, a top-$k$ query needs to be performed on $\theta_l$. Next, the smallest enclosing top-$k'_{\theta_u}$ needs to be computed such that top-$k_{\theta_l} \subseteq$ top-$k'_{\theta_u}$. The points on top-$k'_{\theta_u}$ can now be projected based on $\theta_q$ to compute the answer set (Alg. 4). Note that to index multiple angles, it is not necessary to build as many index structures. Since the information required to perform a range search is independent of the projection angles, a single index structure containing the bounds on llp, lrp, rlp, and rup for each of the indexed angles is enough. More specifically, at each non-leaf node, we store a hashmap. The keys of the hashmap consist of the indexed angles, and each key points to a bucket containing the bounds on the projections for that angle. With the addition of information on multiple angles, the storage cost of the index structure is $O(n + m\frac{n-1}{b-1})$ where $m$ is the number of indexed angles.

Alg. 4 outlines the procedure to compute the top-$k$ answer set on a non-indexed angle. Given an index structure, first, the two consecutive indexed angles are found between which the query angle lies (lines 1-2). Next, the top-$k$ is computed on the smaller of the two indexed angles, $\theta_l$ (line 3). Once top-$k_{\theta_l}$ is computed, top-$k'\theta_u$ is populated by computing top-1 repeatedly on angle $\theta_u$ till all elements in top-$k_{\theta_l}$ are fetched. The score at $\theta_q$ is computed for each of the points fetched in the while loop and maintained in a priority queue (lines 4-8). Finally, the answer set is returned (line 9). Note, the *separating path* is computed only once while computing top-$k_{\theta_l}$ since it is independent of the projection angles.

**Choosing angles to index:** Since a non-indexed angle needs to be bounded within two indexed angles, to cover the entire range of query angles, $0°$ and $90°$ are the two recommended angles. The choice of additional angles to index can be guided by two factors: domain knowledge/history of previous queries, and the main-memory budget. Based on the memory budget, the number of angles to index can first be computed. Next, if the distribution of query projection angles is available, then angles can be indexed based on samples drawn (without replacement) from that distribution. Otherwise, a uniform distribution of angles between $0°$ and $90°$ can be chosen. As shown later in Section 6, an index structure on five angles chosen uniformly is highly efficient in answering the proposed queries. Furthermore, choosing angles from the uniform distribution represents the worst case scenario. Availability of more information can only improve the performance.

## 5. EXTENSION TO HIGHER DIMENSIONS

In this section, we generalize the top-$k$ algorithm to higher dimensions. For both top-1 and top-$k$ queries, the core of the developed algorithms is based on projections. In the 2D space, projections take the shape of a line, and thus the index structures concentrate on indexing line segments. However, in dimensions higher than two, projections take the form of hyperplanes making the problem much harder. Thus, to keep the problem tractable for higher dimensions, we first divide the problem into 2D and 1D subproblems and then solve them individually. Next, we aggregate the subproblems to produce the final answer set.

Eqn. 3 defines the SD-score for any number of dimensions. Recall, $\mathbb{D}$ contains the dimensions which are desired to be distant from the query (repulsive), whereas $\mathbb{S}$ represents the dimensions desired to be similar (attractive). Next, we define subsets $\mathbb{M} \subset \mathbb{D}$ and $\mathbb{N} \subset \mathbb{S}$ where $|\mathbb{M}| = |\mathbb{N}| = min(|\mathbb{D}|, |\mathbb{S}|)$. Further, a bijective function $f : \mathbb{M} \to \mathbb{N}$ is defined that allows us to map each repulsive dimension in $\mathbb{M}$ to an attractive dimension in $\mathbb{N}$. Based on this pairing, the problem is divided into 2D and 1D subproblems. More specifically, Eqn. 3 is reexpressed as the following.

$$\text{SD-score(p,q)} = \left( \sum_{i \in \mathbb{M}, j=f(i)} \alpha_i |q_i - p_i| - \beta_j |q_j - p_j| \right)$$
$$+ \left( \sum_{i \in (\mathbb{D} \setminus \mathbb{M})} \alpha_i |q_i - p_i| \right) - \left( \sum_{j \in (\mathbb{S} \setminus \mathbb{N})} \beta_j |q_j - p_j| \right) \forall p \in \mathbb{P}$$
(10)

As evident from Eqn. 10, each summation corresponds to a subproblem. The first summation in Eqn. 10 involves two dimensions and is solved using the top-$k$ index structure developed in Section 4. The other two subproblems, corresponding to the second and third summations, involve dimensions that need to be solved individually. The proposed formulation maximizes the number of 2D subproblems and consequently, minimizes the total number of subproblems. Eqn. 10 can be completely reduced to 2D subproblems if the cardinalities of $\mathbb{D}$ and $\mathbb{S}$ are same. Otherwise, there will always be some dimension that cannot be paired up to take advantage of the top-$k$ index structure. As the value of $||\mathbb{D}| - |\mathbb{S}||$ increases, SD-score reduces to the familiar top-$k$ similar or distant queries. A high difference in the cardinalities allows less opportunity to employ the developed index structure for better pruning of the search space.

EXAMPLE: *Consider the database of advertisement publishers with attributes $\mathbb{D} = \{Price\}$, and $\mathbb{S} = \{HitRate, Coverage\}$ shown in Figure 6. A possible strategy would be to divide the problem into two subproblems. The first is a 2D subproblem by pairing 'Price' with 'Hit Rate', and the second is a 1D subproblem on 'Coverage'.*

We now discuss how each of the subproblems can be solved and then aggregated to compute the final answer. As already mentioned, the 2D subproblems can be solved using the index structure developed in Section 4. We therefore concentrate on how the 1D subproblems can be solved efficiently. Solving on a single dimension $d$ is performed using a bidirectional search. As part of pre-processing, each dimension is maintained in a sorted container. For a given query $q$, two pointers are maintained to track candidates for top-$k$. If dimension $d$ is desired to be distant, then the pointers are initialized to data points corresponding to the first and last elements in the dimension. On the other hand, if $d$ is desired to be similar, then a binary search is performed with $q_d$ on the selected dimension, where $q_d$ represents the value of $q$ in dimension $d$. If position $pos$ is returned from the binary search then, the pointers are initialized to points corresponding to the elements in position $pos$ and $pos - 1$ in dimension $d$. In essence, after initialization, the pointers point to those two elements that have a chance of being the top-1 point for the subproblem involving dimension $d$. During query-time on dimension $d$, the pointers of $d$ are used to fetch the two candidates and the better candidate is returned as the answer. Further, the pointer corresponding to the answer is updated to the immediate "*unexplored*" neighbor.

EXAMPLE: *To solve the 1D problem on the database shown in Figure 6 for the dimension on 'Coverage', first, it is reordered in the shown sorted container. For the given query, a binary search is performed with its 'Coverage' value of 75. As a result of this search, the two pointers are assigned to $< B, 80 >$ and $< C, 68 >$. Next, to determine the most similar data point on 'Coverage', a comparison is performed only between B and C. Once B is returned, the pointer to B is updated to D.*

The top-$k$ answer set is computed in an iterative manner. At each iteration, the top point is fetched for each of the subproblems. For each of those points, its score against the query is calculated and added to a priority queue of size $k$. Further, a threshold is calculated which plugs the scores from each of the individual subproblems into Eqn. 10 based on the type. The threshold provides an upper bound on the score of any point that has not been explored yet. Thus, the algorithm stops iterating if the $k_{th}$ element in the priority queue has a score higher than or equal to the threshold. The stopping criterion is the same as in Threshold Algorithm [8] and is guaranteed to return an optimal solution.

EXAMPLE: *Expanding on the previous two examples, to compute the top-k answer set on the publishers' database, in the first iteration, the highest scoring points for each of the subproblems, a 2D subproblem on 'Price' and 'Hit Rate' and a 1D problem on dimension 'Coverage', are fetched. Assume the weighting parameters are all equal to 1. Thus, data points A and B are returned as the answers for the 2D and 1D subproblems with scores of 90 and 5 respectively. Therefore, the threshold is $90 - 5 = 85$ since the 2D subproblem corresponds to the first summation in Eqn. 10 and the 1D problem corresponds to the third summation. Further, the scores for A and B, 40 and 45 respectively, are calculated (on the entire query rather than just the subproblems) and added to the priority queue. Similarly, in the next iteration, the second highest scoring points for each of the two subproblems are fetched (C for both subproblems), the threshold is updated to 68, and the points are inserted into the priority queue. The iteration stops when the $k_{th}$ element in the priority queue has a score higher than or equal to the threshold. If $k = 1$, then the computation would stop at the second iteration since SDscore of C is same as the threshold.*

Although the thresholding technique is similar to TA, the key difference lies in the granularity of the subproblems. In TA, and many other top-$k$ techniques [7, 8, 19], each subproblem corresponds to a single dimension. However, in our approach, each subproblem is composed of two dimensions. As a result, a high performance gain is achieved. The impact of the granularity of the subproblems is most prominently evident in scalability against dimension. It is important to note that scalability against dimension for top-$k$ queries is a particularly hard problem. In [18], the authors show that some of the best-performing techniques [3, 10, 18] suffer from scalability bottlenecks above dimensions of size 3. Thus, achieving close to optimal performance on two-dimensional datasets provides a significant boost in raising the dimensionality bar for top-$k$ queries. We quantify the details of the performance gains in Section 6.

| ID | Price | Hit Rate | Coverage | Coverage | Price | Hit Rate | Coverage |
|---|---|---|---|---|---|---|---|
| A | 10 | 40 | 25 | <A, 25> | 150 | 90 | 75 |
| B | 100 | 90 | 80 | <E, 50> | Query | | |
| C | 70 | 85 | 68 | <C, 68> | | | |
| D | 60 | 70 | 85 | <B, 80> | | | |
| E | 90 | 85 | 50 | <D, 85> | | | |
| Database | | | | Sorted Container | | | |

**Figure 6: A sample database and query involving 3 dimensions.**

## 6. EXPERIMENTS

In this section, we report the experimental results that validate the efficiency of the proposed techniques and highlight the applicability on a real dataset.

### 6.1 Experimental Setup

For a thorough evaluation, we use synthetic datasets of up to ten million points generated from uniform, correlated and anti-correlated distributions, and a real dataset of chemical molecules. The index structure for top-$k$ is built on 5 angles distributed uniformly across $90°$: 0, 23, 45, 67, 90. We choose the $\alpha$ and $\beta$ parameters from a uniform distribution between 0 and 1. Unless specifically mentioned, $k$ is set to 5. All experiments are performed on 100 randomly selected points from a uniform distribution.

For benchmarking, we choose sequential scan, and main-memory based adapted versions of TA [8], Branch-and-Bound Processing of Ranked Queries (BRS) [15], and Progressive Exploration (PE) [19]. Note that both BRS and PE were originally designed for disk-resident index structures. We use them for the benchmarking studies since no other main-memory based technique exists that is able to handle non-monotonic functions.

To adapt TA for the proposed class of functions, an ordered list of the data points is maintained for each dimension. Given a query, a binary search is performed to fetch the farthest point on each of the repulsive dimensions and the closest points on the attractive dimensions. The pruning threshold is computed based on the points fetched. To adapt BRS, an in-memory R*-tree is constructed. The node capacity is selected by optimizing the querying performance on a database of 10000 points drawn from a uniform distribution. Based on this optimization, the selected node capacities are 28, 16, 12, and 9 for dimensions 2, 4, 6 and 8 respectively. Given a query, the space is divided into regions such that in each region, the scoring function is either monotonically decreasing or increasing with each dimension. Using the algorithm outlined for constrained top-$k$ queries in BRS, top-$k$ queries are performed simultaneously on each region and the scores are maintained in a single priority queue to compute the answer set. All algorithms are implemented in Java using SUN JDK 1.6.0. The experiments are performed on a 3.2GHz, 4GB memory PC running Debian Linux 4.0.

### 6.2 Quantitative Analysis

First, we benchmark the performance of the proposed index structure in the multi-dimensional case. Figures 7a-7c demonstrate the growth rate of querying time against dataset size for 6-dimensional points. Three dimensions each are chosen for distance and similarity. Clearly, SD-Index top-$k$ displays the lowest growth rate across all three distributions. Although, both TA and SD-Index top-$k$ are based on aggregation of subproblems to compute the final answer set, each subproblem for SD-Index comprises of two dimensions, whereas TA treats each dimension as a subproblem. Due to this basic difference in granularity, SD-Index achieves a better performance. While BRS performs better than TA on the uniformly distributed dataset, TA achieves a superior performance on the correlated and anti-correlated datasets. The bounds computed from the minimum bounding rectangles (MBR) in the uniformly distributed dataset are tighter than the bounds in the correlated and anti-correlated datasets. As a result, a higher pruning is achieved. Compared to sequential scan, SD-Index performs better by an order of magnitude. The performance of PE is similar to sequential scan on 6-dimensional datasets.

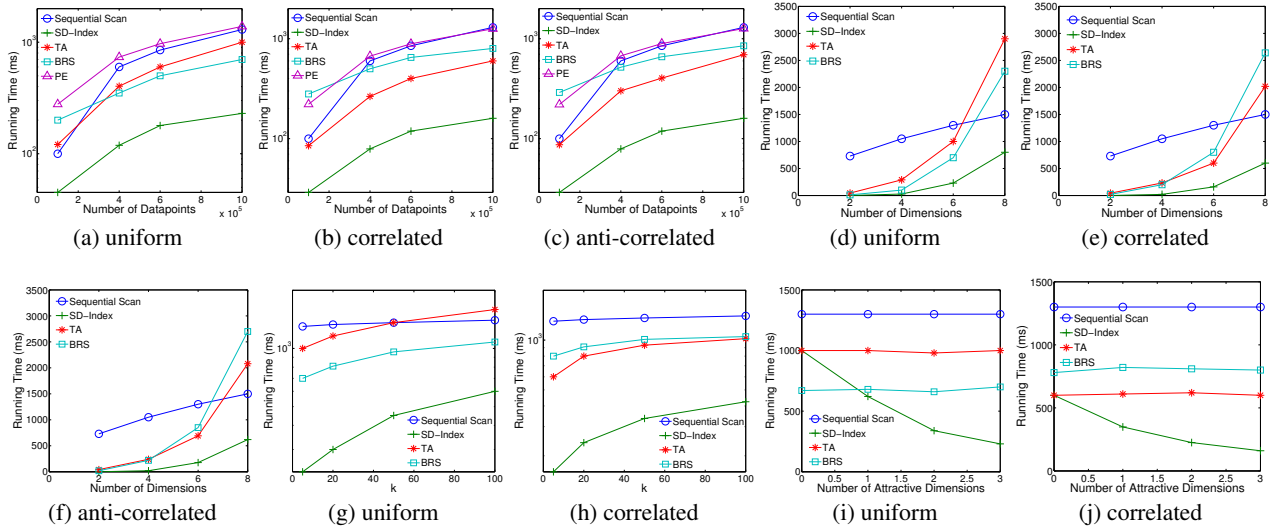

**Figure 7: Growth rate of querying time against dataset size (a-c), number of dimensions (d-f), $k$ (g-h), and number of attractive dimensions (i-j) on uniform, correlated and anti-correlated distributions for SD-Index top-$k$, BRS [15], TA [8], and PE [19].**

In Figures 7d-7f, we analyze the scalability of SD-Index top-$k$ against dimension size. Due to the significantly weaker performance of PE compared to the other methods, we exclude the technique from the remaining benchmarking studies on querying costs. In the top-$k$ setting, well-known techniques have been shown to suffer at dimensions above 3 [18]. Further, none of the recent top-$k$ techniques [15,17,19,20] have analyzed datasets with dimensions above 8. However, even under this context, SD-Index achieves a superior performance at all dimension sizes. As can be seen, both TA and BRS start performing worse than sequential scan at dimensions above 6. As observed in most index structures, sequential scan performs best in terms of scalability. BRS, which is based on hierarchical index structures, suffers at higher dimensions due to the inherent issue of curse of dimensionality [2]. The bounds derived from the MBRs for BRS get progressively looser with higher dimensions. On the other hand, for both SD-Index and TA, more iterations need to be performed for a given $k$ at higher dimensions. However, since each subproblem in SD-Index comprises of two dimensions, the number of iterations is significantly less than TA. This results in superior scalability against dimension when compared to TA, and in general, against the approach of considering each dimension as a subproblem for solving top-$k$ queries [7,8,19].

Figures 7g-7h analyze the growth rate of the querying time against $k$ on a 6-dimensional dataset as $k$ is varied from 5 to 100. We exclude the plot for anti-correlated dataset since the growth rate is similar to the correlated database. As can be seen, SD-Index top-$k$ outperforms sequential scan, BRS, and TA for all values of $k$.

Figures 7i-7j analyze the performance of SD-Index top-$k$ when the number of attractive dimensions is varied. We vary the number of attractive dimensions from 0 to 3 to investigate all possible pairing scenarios. As can be seen, SD-Index top-$k$ achieves a superior performance at all scenarios except when the number of attractive (or repulsive) dimensions is 0. In this setting, SD-Index top-$k$ degenerates into the adapted version of TA and as a result, the performance suffers. It is important to note however, that the basic assumption in the proposed problem is the existence of at least one attractive and repulsive dimensions. Otherwise, the problem translates to a simple distance or similarity query.

Figure 8a analyzes how updates affect the performance of the SD-Index top-$k$ index structure on points drawn from uniform and correlated distributions. We do not analyze the SD-Index top-1 index since its performance is invariant to updates. To analyze the growth rate of querying time with index updates, we deleted and inserted an equal number of randomly selected points to maintain the same index size. Next, we investigated the effect of the updates on the querying cost. SD-Index shows the time without updates, and SD-Index* depicts the querying time after updates. Note that an $x$-axis value of 1000 represents 1000 deletes and 1000 inserts resulting in a total of 2000 updates. As can be seen, the growth rate of the querying time is minimal. Figure 8b demonstrates the growth rate of insertion cost with database size on the four shown index structures. As can be seen, SD-Index top-1 achieves the fastest insertion cost. Although SD-Index top-1 has a worst case insertion cost of $O(n)$, in the average case, it is much faster. Recall that SD-Index top-1 only stores points that can be in the answer set. Since majority of the points do not satisfy this criteria, only an $O(\log m)$ cost is imparted, where $m$ is the number of indexed points. Furthermore, even if the point being inserted can possibly be in the answer set, an $O(n)$ cost is required only when $m = n$, and projections of the point being inserted do not intersect with projections of any of the indexed points. Deletions show a similar behavior.

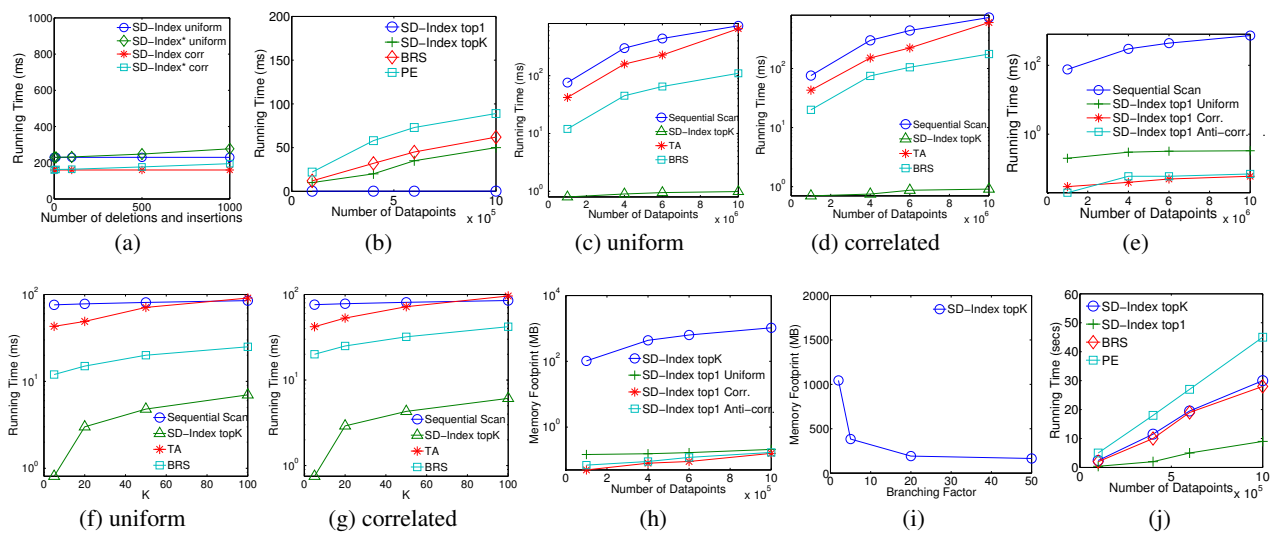

Figure 8: (a) Growth rate of querying cost with updates. (b) Insertion Cost Vs Dataset Size. (c-g) Experiments on 2D data: Querying time Vs. Dataset Size for SD-Index top-$k$ (c-d), top-$1$ (e), Querying Time Vs $k$ (f-g). (h-i) Growth rate of memory footprint against dataset size (h) and branching factor (i). (j) Index Construction Time Vs Dataset Size.

The above results highlight the superiority of SD-Index over TA, BRS, PE and sequential scan in the multi-dimensional case. Now, we verify the performance on 2-dimensional points to demonstrate the performance gain achieved on each of the subproblems.

Figures 8c-8d show the performance of SD-Index top-$k$ against dataset size on 2-dimensional points. We omit the anti-correlated distribution since the result is similar to the performance on the correlated distribution. As can be seen, SD-Index top-$k$ performs better than TA and BRS by more than two and one orders of magnitude respectively. Furthermore, the result shows that due to the significant performance gap between BRS and SD-Index in the 2D case, for high-dimensional datasets, even if BRS is adapted in a manner similar to the proposed strategy of 2D subproblems, SD-Index is still likely to achieve a better performance. Figure 8e repeats the same experiment on the top-1 index structure. Similar to the result in the top-$k$ setting, a performance gain of more than two orders of magnitude over sequential scan is observed. Among the data distributions, a better performance is achieved in the correlated and anti-correlated distributions due to their smaller index sizes. Overall, both the top-1 and top-$k$ index structures display a sub-linear growth rate, which is consistent with the theoretical bounds.

Figures 8f-8g show the growth rate of querying time against $k$ in a 2-dimensional dataset of 10 million points. As can be seen, SD-Index top-$k$ outperforms both BRS and TA by a significant margin. This result also highlights the speed-up that is achieved when a subproblem is composed of two dimensions. Due to this crucial difference in the computation costs of the subproblems, when scaled to higher dimensions or larger $k$, SD-Index achieves a better performance.

Next, we analyze the memory footprint and construction times of the index structures. Figure 8h depicts the growth rate of the memory footprint on a 6-dimensional dataset. Since the index size of SD-Index top-1 is dependent on the data distribution, we investigate the growth rate across all three distributions. As expected, the memory footprint of SD-Index top-1 is much less than the top-$k$ version. This result showcases the advantage of SD-Index top-1 over top-$k$ when $k$ is known apriori. For top-1, datasets drawn from correlated and anti-correlated distributions require less memory since points located at the corners dominate points located between them. As a result, a large number of points can be discarded from the index structure. Figure 8i analyzes the storage cost against the branching factor for SD-Index top-$k$. Figure 8j demonstrates the growth rate of the index construction time against dataset size. As expected, SD-Index top-1 has the fastest index construction time. BRS achieves a marginally faster index construction time than SD-Index top-$k$.

## 6.3 Qualitative Analysis

In this section, we examine the quality of the top-$k$ results on a real dataset of chemical molecules and demonstrate its application in data analysis. For this purpose, we choose the ChEMBL [2] dataset that contains 428,913 bioactive "drug-like" molecules along with calculated properties such as drug-likeness score, logP value, molecular weight (*MW*), etc. Drug-likeness is a concept used in drug design to estimate how "drug-like" a prospective compound is. A good drug should show good availability, low toxicity and high potency. The well-established Lipinski's *rule-of-five* [11] specifies four filters to screen drug-like molecules. Among them, one of the filters states that for a molecule to be drug-like, its MW has to be below 500. For this experiment, we search for exceptions to this rule and check whether those molecules follow any pattern that makes them drug-like in spite of being overweight.

To find exceptions to Lipinski's rule, we queried on a molecule with a high drug-likeness score of 11 and low MW of 250 for similarity on drug-likeness and distance on MW. As a reference point, the highest drug-likeness score and lowest MW in the dataset are 14.22 and 12.01 respectively. The rationale behind the query is to check if overweight drug-like molecules exist. Table 1 summarizes the result. The first row presents the overall average on the selected features, whereas the next four rows present the average for the corresponding values of $k$. Along with the drug-likeness score and MW, Table 1 also presents the polar surface area (abbrv. PSA).

---
[2] http://www.ebi.ac.uk/chembldb/

Table 1: Statistics on top-$k$ results

| Description | Drug-likeness score | *MW* | *PSA* |
|---|---|---|---|
| Overall Average | 8.94 | 422.6 | 112.14 |
| $k$=10 | 9.87 | 938.67 | 27.73 |
| $k$=50 | 9.47 | 897.5 | 42.17 |
| $k$=100 | 9.18 | 877.79 | 42.23 |
| $k$=200 | 9.14 | 824.24 | 47.46 |

Table 1 indicates that molecules with high MW can also be druglike. For all four values of $k$, the average drug-likeness score is more than the overall average, in spite of the average MW in the top-$k$ sets being twice or more of the overall average. A more interesting pattern however, is observed on the PSAs of the molecules. PSA is defined as the sum of the surfaces of polar atoms, usually oxygens, nitrogens and attached hydrogens, in a molecule. As shown in Table 1, PSAs of the top-$k$ molecules are much smaller than the global average. Interestingly, a low PSA has been shown to correlate with human intestinal absorption, and blood-brain barrier penetration making them a good indicator of drug-likeness [16].

The result indicates one of the scenarios where a molecule should not be discarded because of high MW and thereby highlighting an application of SD-query. Traditional distance or similarity based top-$k$ queries fail to identify these molecules due to the limitation of the scoring function. On the other hand, SD-query takes a query-centric approach that is capable of producing a more interesting answer set and thus allowing a deeper analysis of a given dataset.

## 7. RELATED WORK

Top-$k$ queries have been an active research area in the database community. Fagin [7] first introduced the rank aggregation problem for multimedia databases. Since then, an array of techniques has been developed to index top-$k$ queries [3, 4, 8–10, 12, 18, 20]. In the first category, the methods sort each dimension and compute the answer set by making parallel access across dimensions. In the layer-based category, data points are divided into layers to derive an ordering. Points in a layer of higher precedence get inspected before points in a layer of lower precedence. In the view-based category, materialized views are used to answer top-$k$ queries. Unfortunately, none of the above techniques are capable of handling non-monotonic functions.

Among existing techniques, only PE [19] and BRS [15] consider non-monotonic functions. Both [19] and [15] assume datasets to be indexed by disk-resident hierarchical indices and then develop strategies for scoring functions without the assumption of monotonicity. BRS assumes data points to be indexed by a hierarchical index structure such as R-tree, and then computes bounds on the MBRs on any given scoring function to determine whether to further explore its child nodes. PE on the other hand assumes each attribute to be indexed by a hierarchical index structure. Given a query function, it proceeds by exploring the joint space of the indices and computes bounds on whether to explore further. PE defines a special class of *semi-monotone* functions and develops effective pruning strategies for that class.

The key property of both BRS and PE is that the same index structures can be employed to answer a wide range of scoring functions including the proposed class of linear scoring functions. Certainly, BRS and PE are able to handle a wider range of functions than SD-Index. However, for the proposed class of scoring functions, SD-Index achieves a superior performance due to the pre-computation based strategy of indexing the isolines. Furthermore, both BRS and PE are optimized for disk-based index structures, whereas we focus on the memory-resident case.

## 8. CONCLUSION

In this paper, we formulated a novel top-$k$ query on a scoring function that combines the idea of repulsive and attractive dimensions. We developed two index structures for answering top-1 and top-$k$ queries in a scalable manner. Our unique strategy of indexing the isolines of the scoring function achieved a performance gain of one to two orders of magnitude over existing techniques. The application of the proposed class of scoring functions is highlighted in the qualitative analysis of a molecular dataset to identify "drug-like" molecules that deviate from an established rule. For future work, we plan to investigate two primary issues. First, for higher dimensions, the isolines take the form of hyperplanes which are much harder to index. Thus, we want to study whether an index structure can be developed to index hyperplanes directly. Second, while scaling to higher dimensions, the mapping between attractive and repulsive dimensions is currently performed in an arbitrary manner. We plan to analyze whether a more focused strategy can be developed in defining the mapping function since a better mapping would lead to more effective pruning of the search space.

**Acknowledgments:** We thank the anonymous reviewers for their helpful comments. This work was supported by NSF Grant IIS-0917149.